\documentclass{aa}  

\usepackage{graphicx}
\usepackage{txfonts}
\begin{document}

   \title{Improved spectral descriptions of planetary nebulae central stars}

   \author{W. A. Weidmann\inst{1}\fnmsep\thanks{Member of Carrera 
del Investigador CONICET, Argentina.},
          R. H. M\'endez\inst{2}
          \and
          R. Gamen\inst{3}\fnmsep\thanks{Member of Carrera del 
Investigador CONICET, Argentina.}   
          }

\institute{Observatorio Astron\'omico C\'ordoba, Universidad Nacional de 
C\'ordoba, Argentina\\
            \email{walter@oac.uncor.edu}
         \and
Institute for Astronomy, University of Hawaii, 2680 Woodlawn Drive, 
HI 96822 Honolulu, USA
         \and
Instituto de Astrof\'isica de La Plata, CCT La Plata-CONICET, 
Facultad de Ciencias Astron\'omicas y Geof\'isicas
Universidad Nacional de La Plata, Argentina\\
            \email{rgamen@fcaglp.unlp.edu.ar}
}


 
  \abstract
{At least 492 central stars of Galactic planetary nebulae 
(CSPNs) have been assigned spectral types. 
Since many CSPNs are faint, these classification efforts are frequently 
made at low spectral resolution. 
However, the stellar Balmer absorption 
lines are contaminated with nebular emission; therefore in many cases a 
low-resolution spectrum does not enable the determination of
 the H abundance in the CSPN photosphere. 
Whether or not the photosphere is H deficient is arguably the most important fact we 
should expect to extract from the CSPN spectrum, and should be the
basis for an adequate spectral classification system. 
} 
   {Our purpose is to provide accurate spectral classifications and
contribute to the knowledge of central stars of 
planetary nebulae and stellar evolution.
}
   {We have obtained and studied higher quality spectra of CSPNs 
    described in the literature as weak emission-line star (WELS).
We provide descriptions of 19 CSPN spectra. These stars had been
   previously classified at low spectral resolution.
   We used medium-resolution spectra
   taken with the Gemini Multi-Object Spectrograph (GMOS). 
We provide spectral types in the Morgan-Keenan (MK) system whenever 
possible.
}
   {Twelve stars in our sample appear to have normal H rich 
photospheric abundances, and five stars remain unclassified. The rest 
(two) are most probably H deficient.
Of all central stars described by other authors 
as WELS, we find that at least 26\% of them are, in fact, 
H rich O stars, and at least 3\% are H deficient. 
This supports the suggestion that the denomination WELS should 
not be taken as a spectral type, because, as a WELS based on low-resolution 
spectra, it
cannot provide enough information about the photospheric H abundance.}
 {}


   \maketitle


\section{Introduction}

Most main-sequence star spectra can be classified in a simple two-dimensional 
system (surface temperature and surface gravity) because they share almost 
the same chemical composition and do not have strong winds. When we need to 
allow for chemical peculiarities and strong stellar winds, the classification
system necessarily becomes more complicated; that is the case for central
stars of planetary nebulae (CSPNs). In particular, it is very important
to decide which CSPNs are H deficient. This has a direct impact on the
theory of stellar evolution because the simplest picture of post-AGB single 
star evolution \citep[e.g.][]{1989IAUS..131..463S} 
requires a departure from the AGB
before the H rich envelope is depleted, 
leading to the expectation that all CSPNs retain 
a normal H rich composition. In contrast, observations clearly indicate that
about 30\% of CSPNs are H deficient 
\citep[e.g.,][]{1991IAUS..145..375M}. The most popular
solution to this problem is the born-again mechanism 
\citep[e.g.,][]{1984ApJ...277..333I}.

Hence, the empirical information about H deficiency is arguably the most important
fact that a spectral classification for CSPNs should provide. Unfortunately,
to extract this information from CSPN spectra is hard because (1) they are 
faint (60\% of the CSPNs listed in the 
SECGPN\footnote{\cite{1992secg.book.....A}} 
have V$>$15.5; for many other stars in that catalogue no apparent magnitude is 
listed); and (2) the CSPN spectra often suffer contamination from strong 
nebular emissions, in particular, the Balmer lines.

Because of the faintness of CSPNs, a lot of work related to spectral 
classification is done at low spectral resolution, but precisely
because of that fact, the resulting ``spectral types'' frequently fail
to establish whether or not the photosphere is H deficient. 
An example of this situation is the introduction 
by \cite{1993A&AS..102..595T} of the spectral description
weak emission line stars (WELS). 
This denomination essentially  means that
the stellar emission lines are not of Wolf-Rayet type.
Because of the contamination with the nebular emission spectrum at 
low spectral resolution, however, it
is not possible to ascertain if the Balmer absorption lines are present.
Some authors \citep[e.g.,][]{2003IAUS..209..235F,2006PASP..118..183W}
have already remarked that WELS, as a group, are quite heterogeneous.

In particular, in complex cases such as CSPNs, we believe that it is important
to avoid unnecessary confusion, and to restrict the terms ``spectral 
type'' and ``spectral classification'' to those cases where the available 
information permits us to decide whether or not the stellar photosphere
is H rich. In other words, we propose to refrain from using WELS 
as a spectral type, because it induces us to think that all the stars so
described have similar spectra, and that may not be the case.

Of course this statement can be tested empirically, which is the 
purpose of this paper. Of at least 492 CSPNs that have been 
spectroscopically described, 72 have been called WELS
\citep{2011A&A...526A...6W}.
We selected a sample of 19 WELS CSPNs (26\% of the total),
and we endeavoured to obtain higher resolution spectra,
which allows us to qualitatively determine if their photospheres are 
H rich or H deficient.

Section 2 describes the spectrograms and reduction procedures, Section 3
explains the classification criteria, and Section 4 presents the spectral
descriptions and classifications. We present and summarize our conclusions 
in Sections 5 and 6.


\section{Observations and data reduction}

We observed nineteen CSPN described as WELS with Gemini 
Multi-Object Spectrographs (GMOS-N and GMOS-S) at the Gemini 
Telescopes North and South, see Table~\ref{sample}. 
We ordered these objects by Galactic longitude.

Our spectrograms were acquired under programs 
GS-2013A-Q-44,
GN-2014A-Q-107 and 
GN-2014B-Q-101
 (PI: W. Weidmann).
We increased our sample with all GMOS CSPN spectra that were 
publicly available in the Gemini database.

For our programs, we selected the B1200 grating (R$\simeq$2300) at long-slit mode
used with a 0.75\arcsec\ slit, or grating B600 (R$\simeq$1700) with a 1\arcsec\ slit.
The selected spectral range (see Table~\ref{sample}) is useful to
identify/classify O-type as well as Wolf-Rayet stars.

We chose the science exposure times as a function 
of the CSPN magnitude \citep{1990secg.book.....A} 
to reach a S/N up to 100 (see Table~\ref{sample}).
The strong nebular emission lines frequently saturated the CCD chip.

To reduce the data, we followed standard procedures using tasks in
the gemini.gmos IRAF package as well as generic IRAF\footnote{IRAF is 
distributed by the National Optical Astronomical Observatories, 
operated by the Association of Universities for Research in Astronomy, 
Inc., under contract to the National Science Foundation of the USA.}
tasks. The reduction procedures included image trimming, flat fielding, 
wavelength calibration, sky subtraction, and relative flux calibration
(the spectrophotometric standard star EG~131 was used for this purpose, see
\citealp{1999PASP..111.1426B}).

The spectra were reduced separately and then combined for the final 1D-spectral extraction.
We rectified the continua, and shifted the spectra  in wavelength to zero radial velocity.
Whenever possible, the contribution from the nebula was subtracted 
by interpolating nebular regions on opposite sides of the CSPN.

\begin{table*}
\caption{GMOS spectra of WELS CSPNs.}             
\label{sample}      
\centering                          
\begin{tabular}{c l l c l c}        
\hline\hline                 
PN G & name & Gemini program & spectral range [\AA] & grating & integration [sec] \\    
\hline    
 000.3$+$12.2 & IC~4634  & GS-2012A-Q-69  & 3767--5977    & B1200 & 1 $\times$ 900    \\ 
 009.6$+$14.8 & NGC~6309 & GS-2011A-Q-65  & 3767--5225    & B1200 & 1 $\times$ 1800   \\ 
 015.4$-$04.5 & M~1-53   & GS-2013A-Q-44  & 4500--5970    & B1200 & 3 $\times$ 390    \\   
 025.8$-$17.9 & NGC~6818 & GS-2011A-Q-65  & 3770--5220    & B1200 & 1 $\times$ 1800   \\ 
 034.6$+$11.8 & NGC~6572 & GN-2014A-Q-107 & 3700--6500    & B600  & 1 $\times$ 1200   \\ 
 038.2$+$12.0 & Cn~3-1   & GN-2014A-Q-107 & 3700--6500    & B600  & 3 $\times$ 800    \\             
 051.9$-$03.8 & M~1-73   & GN-2014B-Q-101 & 3970--5430    & B1200 & 3 $\times$ 450    \\
 054.1$-$12.1 & NGC~6891 & GN-2014A-Q-107 & 3700--6500    & B600  & 3 $\times$ 600    \\
 057.2$-$08.9 & NGC~6879 & GN-2014B-Q-101 & 3970--5430    & B1200 & 3 $\times$ 500    \\
 058.3$-$10.9 & IC~4997  & GN-2014A-Q-107 & 3700--6500    & B600  & 3 $\times$ 1200   \\
 096.4$+$29.9 & NGC~6543 & GN-2014A-Q-107 & 3700--6500    & B600  & 3 $\times$ 200    \\
 159.0$-$15.1 & IC~351   & GN-2014B-Q-101 & 3970--5430    & B1200 & 3 $\times$ 900    \\
 190.3$-$17.7 & J~320    & GN-2014B-Q-101 & 3970--5430    & B1200 & 3 $\times$ 400    \\
 307.2$-$09.0 & He~2-97  & GS-2013A-Q-44  & 4500--5970    & B1200 & 2 $\times$ 210    \\
 312.3$+$10.5 & NGC~5307 & GS-2011A-Q-65  & 3767--5228    & B1200 & 1 $\times$ 900    \\ 
 345.0$-$04.9 & Cn~1-3   & GS-2013A-Q-44  & 4500--5970    & B1200 & 2 $\times$ 340    \\
 348.0$-$13.8 & IC~4699  & GS-2013A-Q-44  & 4500--5970    & B1200 & 3 $\times$ 410    \\
 349.3$-$01.1 & NGC~6337 & GS-2011A-Q-91  & 4000--5520    & B1200 & 4 $\times$ 600    \\ 
 355.7$-$03.5 & H~1-35   & GS-2013A-Q-44  & 4500--5970    & B1200 & 2 $\times$ 760    \\ 
\hline 
\end{tabular}
\end{table*}

\begin{table*}
\caption{Improved spectral classification and key lines in 
medium resolution spectra of WELS CSPNs.
The N.S. column indicates whether or not it was possible to 
subtract the nebular component.
The letters A and E mean that the ion was in absorption and 
emission, respectively. An undetected ion was denoted (-), 
and N/D means no data available in the corresponding spectral range.
The columns labeled ``\ion{N}{v}'', ``\ion{C}{iv}'' and ``\ion{O}{v}''
indicate the doublet at 4603-19\AA, the doublet at 5801-11\AA\, and 
5114\AA\, respectively. 
}
\label{medicion}      
\centering                          
\begin{tabular}{c l c c c c c c c c c}        
\hline\hline                 
 PN G         & name     & S.T.              & 4542 & 4686 & 5412 & H$\beta$ & \ion{N}{v} & \ion{C}{iv} & \ion{O}{v} &  N.S.\\
\hline    
\object{000.3$+$12.2} & IC~4634  & O3 I f*           & A    & AE     & A   & A  & A         & E   & A & y  \\  
\object{009.6$+$14.8} & NGC~6309 & O(He)             & A    & A      & A?  & A  & E         & E?  & - & y  \\ 
\object{015.4$-$04.5} & M~1-53   & O3 I f            & A    & E      & A   & A  & A         & E   & A & y  \\    
\object{025.8$-$17.9} & NGC~6818 & not classif       & A?   & E      & A?  & A? & -         & E   & - & y  \\  
\object{034.6$+$11.8} & NGC~6572 & Of-WR(H)          & A    & E      & E   & E  & E         & E   & A & y  \\
\object{038.2$+$12.0} & Cn~3-1   & O7 Ib (f)         & A    & -      & A   & A  & -         & A   & - & n  \\
\object{051.9$-$03.8} & M~1-73   & O3.5 I f          & A    & E      & A   & AE & A         & E   & - & y  \\
\object{054.1$-$12.1} & NGC~6891 & O3 Ib (f*)        & A    & AE     & A   & A  & A         & E   & A & y  \\
\object{057.2$-$08.9} & NGC~6879 & O3f(He)           & A    & E      & A   & A? & A         & E?  & A & n  \\
\object{058.3$-$10.9} & IC~4997  & not classif       & A    & E      & -   & -  & A?        & E   & - & n  \\
\object{096.4$+$29.9} & NGC~6543 & Of-WR(H)          & A?   & E      & E   & E  & E         & E   & A & y  \\
\object{159.0$-$15.1} & IC~351   & Of(H)             & A    & E?     & A   & A  & E         & N/D & A? & n \\
\object{190.3$-$17.7} & J~320    & O3 V ((f))        & A    & A      & A   & A  & A         & N/D & A & y  \\
\object{307.2$-$09.0} & He~2-97  & Of-WR(H)          & AE   & E      & -   & E  & E         & E   & - & n  \\ 
\object{312.3$+$10.5} & NGC~5307 & O3.5 V            & A    & A      & A   & A  & A         & E   & A & y  \\  
\object{345.0$-$04.9} & Cn~1-3   & not classif       & A    & E?     & A   & E? & -         & E   & - & n  \\ 
\object{348.0$-$13.8} & IC~4699  & O3 V ((f))        & A    & A      & A   & A  & A         & E   & A & y  \\ 
\object{349.3$-$01.1} & NGC~6337 & not classif       & E    & E      & E   & E  & E         & E   & - & y  \\  
\object{355.7$-$03.5} & H~1-35   & not classif       & A    & E?     & A?  & E  & E?        & E?  & E? & n \\  
\hline    
\end{tabular}
\tablefoot{References for the \ion{C}{iv} doublet at 5806 \ \AA\
   are \citet{2011A&A...531A.172W},
       \citet{1993A&AS..102..595T}, and
the spectra shown in Figs.~\ref{ST-red1} to \ref{ST-red3}.}
\end{table*}


\section{Spectral classification criteria}

Given the high surface temperatures of CSPNs, we clearly have to deal with 
the earliest spectral types in the MK system. The MK system has recently been 
extended \citep{2011ApJS..193...24S,2014ApJS..211...10S},
building on earlier work by Nolan Walborn. 
See also  \citet{2010ApJ...711L.143W}.
It should be obvious that only some CSPN spectra can be given an MK type. 
The MK system presupposes a normal H rich photospheric composition and low
or, at most, moderate mass-loss rate. If any of these conditions are not met,
each individual spectrum becomes almost unique, and the exercise of spectral
classification becomes much harder.

We decided to proceed in the following way. For H rich stars 
classifiable as O and Of, we tried to provide the best possible 
classification in the system of \citet{2011ApJS..193...24S}.
Note, in particular, the list of qualifiers in their Table~3. 
It should be clear that using the Sota et al. spectral types 
necessarily implies that the star in question is demonstrably H rich. 
The ``luminosity classes'' should instead be interpreted as ``surface gravity classes''; 
since CSPNs are less massive, they are of course 
less luminous than the typical massive O-type stars of Sota et al. 
with similar luminosity classification.

Among CSPNs we also find spectra that are
quite different from any MK standard. In these cases, we prefer to follow
the classification scheme of \citet{1991IAUS..145..375M}. 
In particular, we follow the spectral
type Of-WR(H), indicating a denser stellar wind, with broad emission lines
and blue-shifted profiles for key diagnostic lines, plus essentially normal 
H rich composition and the spectral type O(He), which indicates a H poor 
photosphere with He as the predominant element. Our identification of 
H or He as the most abundant photospheric element is based on the behavior 
of the intensities of even-n and odd-n Pickering \ion{He}{ii} lines; 
the even-n lines blend with the corresponding Balmer lines.

\begin{figure*}
   \centering
   \includegraphics[width=0.99\textwidth]{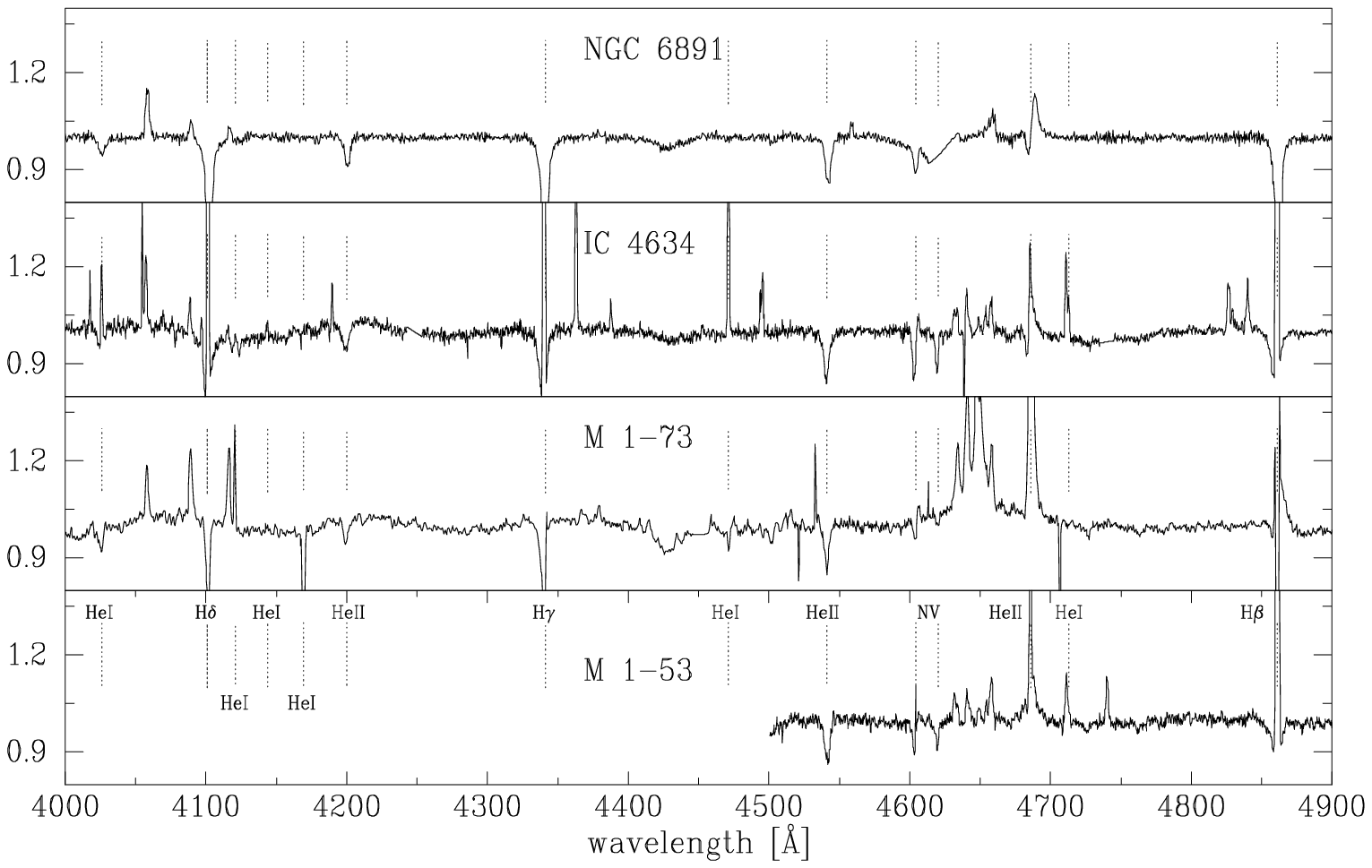}
   \includegraphics[width=0.99\textwidth]{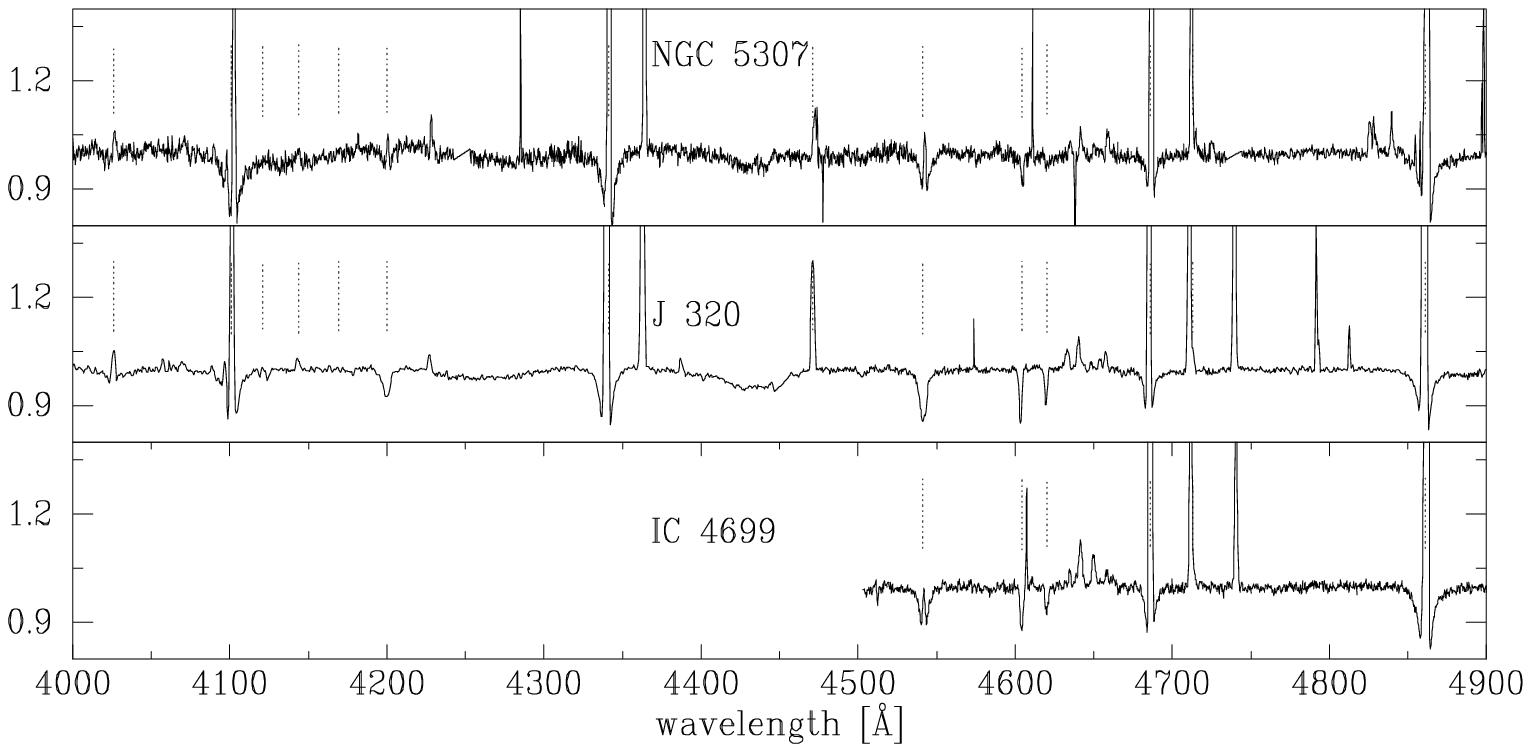}
      \caption[]{Normalized spectra of O-type CSPN grouped
                 according to their spectral classification: O3I and O3V 
                 (see Table~\ref{sample}).
                 The interstellar absorption band at $\lambda$4428 are not indicated.
                 The most important spectral features (absorption or emission) identified are
   H$\beta$, H$\gamma$, and H$\delta$.
   \ion{He}{i} $\lambda$4026, 4121, 4144, 4169, 4471, and 4713.
   \ion{He}{ii} $\lambda$4200, 4541, and 4686. 
   \ion{N}{v} $\lambda$4604-19.
}
         \label{ST-type-0}
   \end{figure*}

\begin{figure*}
   \centering
   \includegraphics[width=0.99\textwidth]{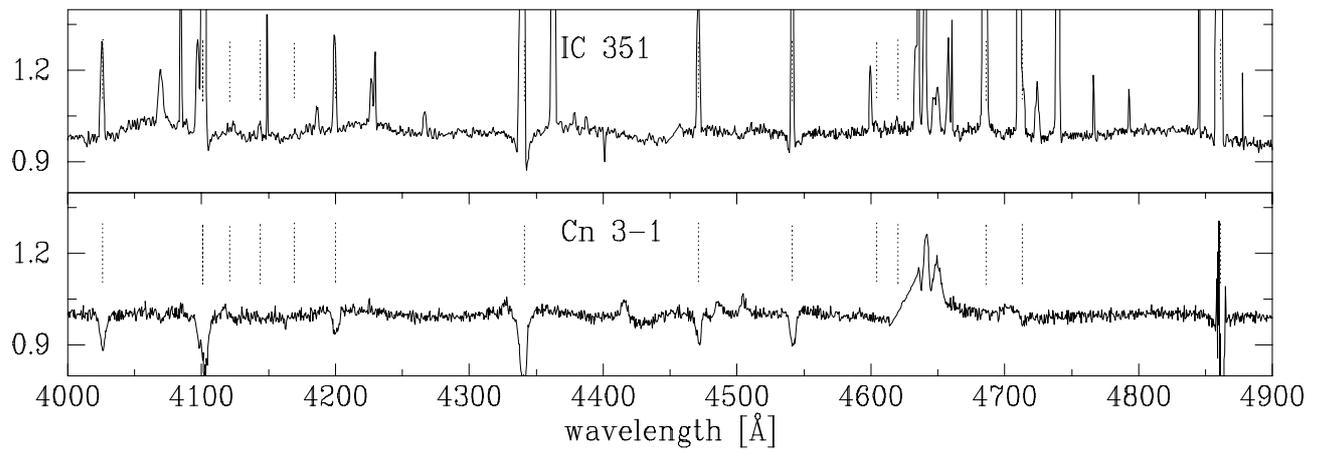}
      \caption[]{Normalized spectra of early-O CSPN. }
         \label{ST-type-1}
   \end{figure*}

\begin{figure*}
   \centering
   \includegraphics[width=0.99\textwidth]{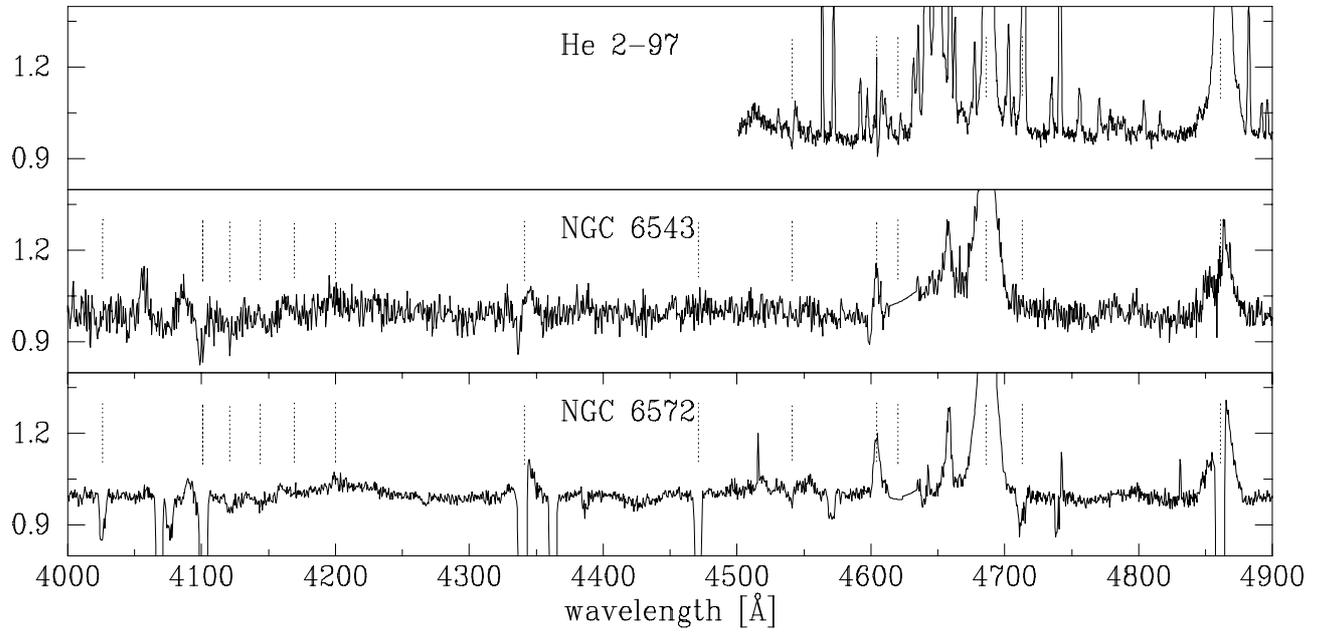}
      \caption[]{Normalized spectra of Of-WR(H) CSPN.
                 Note the broad emission of \ion{He}{ii} at 4686\AA. }
         \label{ST-type-2}
   \end{figure*}

\begin{figure*}
   \centering
   \includegraphics[width=0.99\textwidth]{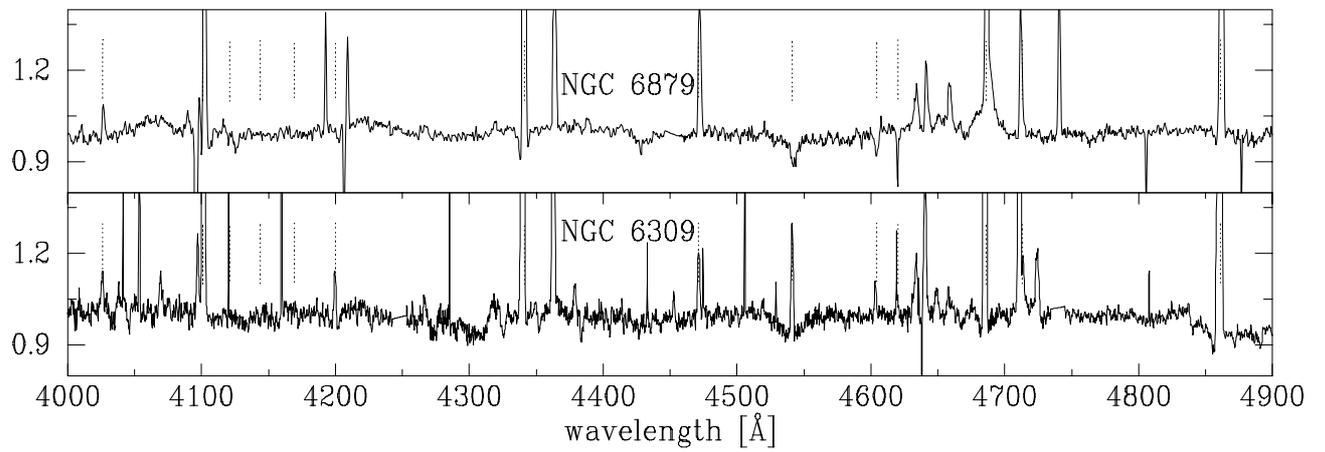}
      \caption[]{Normalized spectra of possible O(He) CSPN. }
         \label{ST-type-8}
   \end{figure*}

\begin{figure*}
   \centering
   \includegraphics[width=0.99\textwidth]{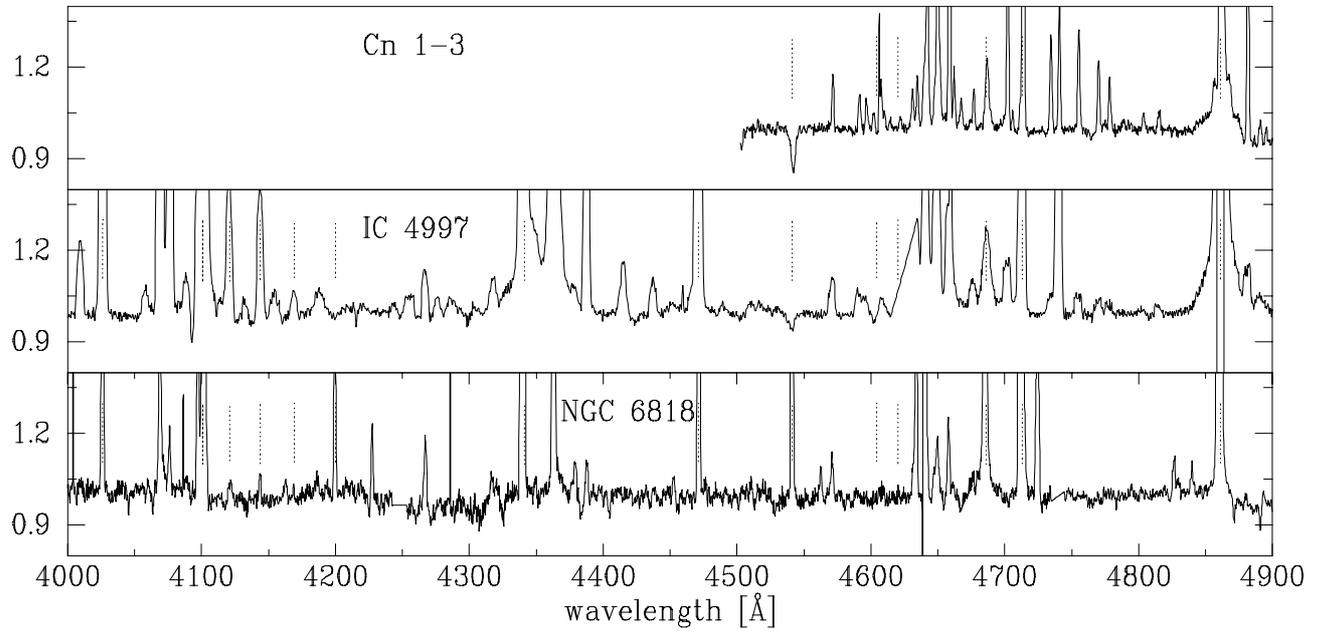}
   \includegraphics[width=0.99\textwidth]{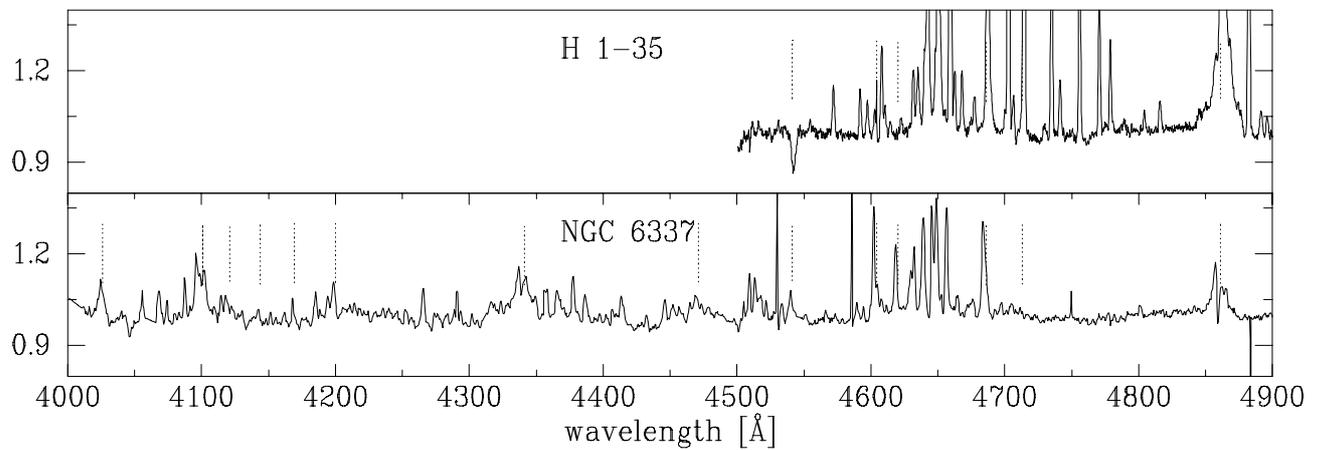}
      \caption[]{CSPN without classification.
                 Note the evident absorption of \ion{He}{ii} in Cn~1-3, IC~4997, and H~1-35.}
         \label{ST-type-3}
   \end{figure*}

The determined spectral types are given in Table~\ref{medicion},
and the spectra are shown in Figs.~\ref{ST-type-0} to \ref{ST-type-3}, 
arranged according to the spectral type.
In addition, we include the red part of the spectra
(Fig.~\ref{ST-red1} to \ref{ST-red3})
to highlight the presence and intensity of ions such as
\ion{O}{v} and \ion{C}{iv}.

In next section, we present individual descriptions for each of 
the central stars we observed. 
In some cases, the strength of the nebular emission lines 
made classification impossible (H abundance undecided). 
In these cases the only solution is higher spectral resolution 
and/or much more efficient nebular subtraction, which can be 
obtained, for example, in excellent seeing conditions using 
a much narrower spectrograph slit.


\section{Stellar descriptions and classifications: notes on individual objects}

\textbf{IC~4634}: Balmer and  \ion{He}{ii} absorption lines clearly visible. 
 The 4686\AA\ emission line shows a P Cygni-type profile, obviously of 
stellar origin. Together with a \ion{N}{iv} 4058\AA\ emission stronger 
than \ion{N}{iii} 4640\AA\ emission, this indicates a qualifier f*.
The \ion{N}{v} absorption doublet is strong, but weaker than \ion{He}{ii}
4541, leading to a spectral type O3.

\textbf{NGC~6309}: \ion{He}{ii} 4686 and 4541 in absorption. Since 4541
appears to be stronger than any absorption at 4340, we believe that this
star is a new example of spectral type O(He) \citep{1991IAUS..145..375M}.

\textbf{M~1-53}: Balmer and \ion{He}{ii} 4541 in absorption. The \ion{N}{v} 
4603-19\AA \ absorption doublet is present, but weaker than 4541, indicating
a spectral type O3. Other lines present: \ion{N}{iii} 4634-40-42\AA, 
\ion{C}{iv} 4658\AA \ emissions, \ion{O}{v} 5114\AA \ absorption.
The presence of \ion{O}{vi} emission at 5290\AA \ further suggests a very 
hot star.

\textbf{NGC~6818}: this nebula's angular size made
a reasonable nebular emission subtraction possible. 
However, the nebular emissions are so strong that no stellar feature 
is clearly visible, with the only exception of \ion{He}{ii} 4686 emission which
we could not classify. Higher spectral resolution will be necessary.

\textbf{NGC~6572}: broad Balmer and \ion{He}{ii} 4686 emissions (most
absorptions seen in our spectrum are an artifact of nebular subtraction).
Note that 4541 is not in emission, therefore this star cannot be classified 
as a Wolf-Rayet.
There may be a stellar absorption of \ion{C}{iii} at 4069\AA. There is
strong \ion{C}{iii} emission at 4658\AA. \ion{O}{vi} emission is visible at 5290\AA.
This star has been classified as Of-WR(H) by 
\citet{1991IAUS..145..375M}.
 A good spectrum longward of 4200\AA \ has been published by 
\citet{1990A&A...229..152M}.

\textbf{Cn~3-1}: Balmer lines in absorption. Both \ion{He}{i} 4471 and
\ion{He}{ii} 4541 clearly seen in absorption; this indicates a rather 
low-temperature star. From the ratio 4471/4541 it is an O7 star.
The \ion{C}{iv} doublet at 5806\AA\ is in absorption, again consistent
with a lower temperature (this doublet changes from absorption into 
emission at effective temperatures around 50,000 K; see Section 5.1 below).
There is absorption of \ion{C}{iii} and \ion{N}{iii}
at 4070 and 4097\AA, respectively,
together with \ion{Si}{iv} at 4089 and 4116\AA. Since there is emission
at \ion{N}{iii} 4634-40-42\AA, and \ion{He}{ii} 4686 is absent, presumably
filled with emission, we adopt a qualifier (f) for this star.
Note the presence of \ion{S}{iv} emission at 4486, 4504\AA. Another nice 
example of this \ion{S}{iv} doublet in a central star spectrum is shown in
Fig.~1 of \cite{2007A&A...467.1253H}.

\textbf{M~1-73}: 
the Balmer profiles are affected by nebular subtraction, but H$\beta$
appears to show a P Cygni profile. Well-defined absorption lines 
of \ion{He}{ii} (at 5412, 4541 and 4200\AA). The \ion{He}{ii} stellar 
feature at 4686 has an equivalent width of 4.3\AA, suggesting 
photospheric emission (the nebular FWHM is 2.2\AA),  
therefore we classify this star as Of.
The absorption lines of \ion{N}{v}  suggest an O3.5 star.

\textbf{NGC~6891}: a reasonably good nebular emission subtraction 
was possible. The Balmer series and \ion{He}{ii} absorptions are
clearly visible. \ion{He}{ii} at 4686\AA \ shows a P Cygni profile.
The \ion{N}{v} absorption at 4603 \AA \ is clearly present (the 
4619 line is not visible because of a CCD gap). We adopt a type O3.
The weakness of \ion{N}{iii} at 4640\AA\ and the prominent emission of 
\ion{N}{iv} at 4058\AA \ suggests a qualifier (f*). 
\ion{Si}{iv} at 4088, 4116 \AA, and \ion{C}{iv} at 4658\AA \ appear in emission.
This CSPN was classified as Of(H) \citep{1991IAUS..145..375M}, essentially
in agreement with our classification.

\textbf{NGC~6879}: compact nebula, very strong nebular emissions.
Since the absorption at \ion{He}{ii} 4541 appears to be slightly
stronger than the absorption at 4340, we believe that this
star may be H deficient, although a higher resolution spectrum would be
desirable. The presence of \ion{N}{v} 4603-19 indicates an early
spectral type. Since \ion{N}{iii} 4634-40-41, and \ion{He}{ii} 4686
are in emission, we adopt a spectral type O3f(He), following 
\citet{1991IAUS..145..375M}. 
\ion{Si}{iv} 4088, 4116 appear to be in emission.

\textbf{IC~4997}: this nebula is spatially unresolved, and the stellar
spectrum is severely contaminated with nebular emissions. We see
absorption lines of  \ion{He}{ii} (4542, 4200 and 5412\AA). The presence
of photospheric H cannot be decided. The emission of \ion{He}{ii} at 
4686\AA\ is wider than nebular lines, suggesting a stellar origin.

\textbf{NGC~6543}: this star has been classified as Of-WR(H) by 
\citet{1991IAUS..145..375M},
indicating that it is H rich. The spectrum is similar to that of 
NGC 6572.  \citet{2008ApJ...681..333G} analyzed 
a high quality spectrum of this star.
Although they did not propose any spectral classification,
they confirmed that the star is not H poor.
Our spectral range allows us to describe
\ion{C}{iii} at  5696 \AA \ and \ion{C}{iv} at 5806 \AA. 
The former is absent, but the \ion{C}{iv} emission is very intense and 
broad, confirming the impression that this star's wind is somewhat
intermediate between Of and Wolf-Rayet, in agreement with  
\citep{1991IAUS..145..375M}. 
Other emission lines we identified are
\ion{N}{iv} at 4058\AA\ and \ion{N}{v} at 4603\AA, \ion{C}{iv} at 4658\AA, 
\ion{He}{ii} at 4686 and 5412\AA, \ion{O}{vi} at 5290\AA.

\textbf{IC~351}: 
the stellar spectrum shows clear evidence of H$\gamma$ 4340 and \ion{He}{ii}
4541, both in absorption. The \ion{He}{i} line at 4471\AA\ is not visible
because of nebular contamination, so we cannot estimate a reliable O subtype. 
This star is probably Of, although we cannot decide if there is any 
\ion{He}{ii} stellar emission at 4686\AA.

\textbf{J~320}: this object shows a clearly defined set of H and 
\ion{He}{ii} absorption lines, which together with strong \ion{N}{v} 4603-19
absorptions and weak \ion{N}{iii} 4634-40-41 emissions is consistent 
with a spectral type O3((f)).

\textbf{He~2-97}: this nebula is not resolved and the nebular lines are quite 
strong, however, part of the H$\beta$ emission appears to be of stellar 
origin. The emissions of \ion{He}{ii} and \ion{C}{iv} at 4686 and 
5801-12\AA\ are wide (FWHM 6.5, 5.5 and 5.9\AA, respectively, while 
nebular lines have a FWHM of 2.0\AA). An interesting feature is the P Cygni 
profile of \ion{He}{ii} at 4541\AA, indicating a dense stellar wind. 
This central star looks similar to those of NGC 6543 and NGC 6572, but 
with stronger \ion{C}{iii} 4650. 
Following \citet{1991IAUS..145..375M}, we have adopted a type Of-WR(H).

\textbf{NGC~5307}: the spectrum shows a clear Balmer series and 
\ion{He}{ii} lines all in absorption, in particular at 4686\AA.
Absorption of \ion{N}{v} at 4603\AA\ is visible.
In addition, weak \ion{C}{iv} emission at 4658\AA\ is visible.
We adopt a spectral type O3.5 V.

\textbf{Cn~1-3}: the only clear absorption we see is \ion{He}{ii} 4541.
The emission at 4686\AA\ is wider than the nebular lines, suggesting 
a stellar component and therefore a spectral type Of. It is not clear if
H$\beta$ has a stellar emission component, and therefore we refrain from 
classifying this star.

\textbf{IC~4699}: the spectrum shows a clear Balmer series and 
\ion{He}{ii} lines all in absorption, in particular at 4686\AA.
Absorptions of \ion{N}{v} at 4603, 19\AA\ are visible.
In addition, weak \ion{N}{iii} emissions at 4634, 40, 41\AA\ are visible.
We adopt a spectral type O3 V ((f)).

\textbf{NGC~6337}: this central star is a binary 
 \citep{2006IAUS..234..421H}, 
detected through photometric variability. 
The emission lines are probably due to the 
irradiated companion \citep{2008AJ....136..323D}.
No spectrum of this star has been published previously.
Our spectrum looks similar to that of the central star of
\object{IPHASXJ194359.5+170901} \citep{2011MNRAS.410.1349C}.
There is no clear absorption line, however, we see a combination of 
absorption and emission lines at the Balmer series. 
This is confirmed by  
spectra taken at CASLEO (El Leoncito Astronomical Complex, San Juan, 
Argentina), with the 215 cm Sahade telescope; these spectra 
display a strong H$\alpha$ absorption. 
This suggests that the brightest star in the binary system 
is not a late-type star, and that it is H rich, but we refrain from 
classifying this star. There is an emission blend near 4650\AA, i.e., 
\ion{N}{v} 4603, 19, \ion{N}{iii} 4634-41\AA, \ion{C}{iii} 4647-51\AA, 
\ion{C}{iv} 4658\AA, and \ion{He}{ii} 4686\AA. This collection of 
emission lines led to describe this star as WELS.

\textbf{H~1-35}: the available section of the spectrum is very similar 
to that of Cn~1-3. There seems to be a stellar H$\beta$ emission component,
and the only absorption feature visible is \ion{He}{ii} 4541. We refrain
from classifying this star.


\section{Discussion and interpretation of our classifications}

The WELS denomination was introduced by \cite{1993A&AS..102..595T}.
It has been used when some of the following stellar emission lines are 
observed: the 4650 group (\ion{N}{iii} 4634 $+$ \ion{N}{iii} 4641 $+$ 
\ion{C}{iii} 4647 $+$ \ion{C}{iii} 4650 $+$ \ion{C}{iv} 4658),
\ion{He}{ii} 4686, and \ion{C}{iv} 5801-11. 
We explained in the 
introduction why we think this denomination should not be used as a 
spectral type. Other authors have expressed similar reservations:
\citet{2012IAUS..283..107M} and 
\citet[subsection~5.3]{2014RMxAA..50..203K}.
We now summarize the result of our empirical test.

Our spectroscopic survey of 19 WELS stars has indeed shown a variety 
of spectral types. Nine CSPNs turn out to be H rich O and Of stars.
Three can be classified as Of-WR(H) stars, in view of their denser stellar 
winds \citep{1991IAUS..145..375M}.
In fact, two of them (NGC 6543 and NGC 6572) had
been previously described as H rich, based on high-resolution spectra, 
before being described as WELS. 
We did not find any Wolf-Rayet star.

Hence, in our sample of 19 CSPNs, we found a total of 12 H rich stars. 
We classified two stars as probably O(He), although higher resolution 
spectra would be desirable to provide better confirmation. Five stars
could not be reliably classified, again requiring a higher spectral 
resolution. We confirm that the denomination WELS means, in practice,
``insufficient spectral information'' to decide if these stars are H rich 
or not. Most of the WELS stars turn out to be H rich.

For completeness, it may be useful to add (see Table~\ref{remedida}) a list of CSPNs,
described as WELS from low-resolution spectra, that are in fact O and
Of H rich stars, previously and/or subsequently classified as such on the 
basis of better spectrograms.

\begin{table*}
\caption{Other O- and Of-type CSPNs described as WELS.}             
\label{remedida}      
\centering                          
\begin{tabular}{c l c c c}        
\hline\hline                 
 PN~G         & name     & S.T.                &  References  & CIV emission ref. \\    
\hline    
 002.0$-$06.2 & M~2-33   & O5f(H)              &  HP2007      & HP2007           \\
 002.1$-$02.2 & M~3-20   & Of                  &  AK1987      & em? GC2009       \\
 003.9$-$14.9 & Hb~7     & O3                  &  GP2001      & no em TA1993     \\
 016.4$-$01.9 & M~1-46   & Of(H)               &  H2003       & no em TA1993     \\
 264.4$-$12.7 & He~2-5   & Of(H)               &  M1991       & N/D              \\  
 315.1$-$13.0 & He~2-131 & O8(f)p              &  MN1979      &  no em TA1993    \\ 
 316.1$+$08.4 & He~2-108 & Of(H)               &  M1991       & PCygni MA2003    \\  
\hline    
\end{tabular}
\tablefoot{
HP2007 \citep{2007A&A...467.1253H},
H2003 \citep{2003IAUS..209..237H},
GP2001 \citep{2001A&A...373..572G},
M1991 \citep{1991IAUS..145..375M},
AK1987 \citep{1987ApJS...65..405A},
MN1979 \citep{1979ApJ...232..496M}.
GC2009 \citep{2009A&A...500.1089G},
TA1993 \citep{1993A&AS..102..595T},
MA2003 \citep{2003AJ....126..887M}
}
\end{table*}


\subsection{The \ion{C}{iv} emission}

A common feature in stars described as WELS is the emission doublet
of \ion{C}{iv} at 5801-11\AA. It is commonly attributed to the CSPN,
but in some cases it may be due to the nebula instead; see the discussion
of the stellar spectrum of NGC~5979 in \citet{2014A&A...570A..26G}.
Clearly, if the PN is spatially unresolved and of high excitation, it
may be hard to decide its origin. A higher spectral resolution would be
required, for example to verify if a double-peaked emission due
to nebular expansion can be resolved.

However, in many CSPNs this \ion{C}{iv} doublet is clearly of stellar
origin. We compare now this \ion{C}{iv} emission  with Pop~I stars. 
According to \citet{1974ApJ...187..539C}, the \ion{C}{iv} 
doublet is not seen in emission, even in the most extreme Of stars. 
\citet{1995bces.book.....J} agrees, also noting that O-type stars 
show this doublet in absorption, with a maximum around O7. 
\citet{2001ASPC..242..217W} recognizes that the emission of \ion{C}{iv} at 5806\AA\
has not received too much attention in the literature.
In addition, \ion{C}{iv} emission is not reported in sdO stars
\citep{2013A&A...551A..31D}.

\citet{2000PASP..112.1446W} have shown that these \ion{C}{iv} 
lines are visible in emission in the spectrum of O3~If* type stars,
almost neutral in O4~If, and as pure absorption lines in O5~If type star.
Moreover, they are also in emission in the spectrum of the O2~I star HD~93129A.

In summary, these narrow emission lines are common in the O-type stars 
that are nuclei of planetary nebulae but not in those of Population~I.
The difference could be a surface temperature effect.
Indeed, CSPNs with T$_{eff}$<45,000 K always display \ion{C}{iv} at 5806\AA\
in absorption (e.g., He~2-138, M~1-26, Tc~1, IC~418).
On the other hand, CSPNs with T$_{eff}$>50,000 K usually present 5801-11 
in emission (e.g., NGC~1535, NGC~3242, NGC~4361, NGC~7009).
As we described, in Pop~I O-type stars these emissions are
only detected in the spectra of O2~I and O3~If*. 
In these cases \citet{2005A&A...436.1049M} determined a 
T$_{eff}$<50,000 K, so perhaps the transition temperature from 
absorption to emission is somewhat lower than 50,000 K.


\section{Summary of conclusions}

The number of CSPNs described as WELS in the literature is 72.
Our spectroscopic survey of 19 WELS stars has shown a variety
of spectral types. Of these, we determine that 12 have H rich 
atmospheres, with different wind densities. 
In five cases we could not decide (however, they are not [WR]).
Two cases seem to be H-deficient, although even better spectra
would be desirable to confirm this.

We find no reason to assume that the WELS denomination is
predominantly associated with H deficiency, and our empirical test
indicates that the use of WELS as a spectral type should be discouraged. 
The WELS denomination should be restricted to mean
``insufficient spectral information'' to decide if these stars are
H rich or not. Most of the WELS stars, in fact, turn out to be H rich.

A by-product of this study is that we have found several CSPN spectra
with strong, well-defined metal absorption and emission lines that
would be suitable targets in a search for spectroscopic binaries.


\begin{figure*}
   \centering
   \includegraphics[width=0.99\textwidth]{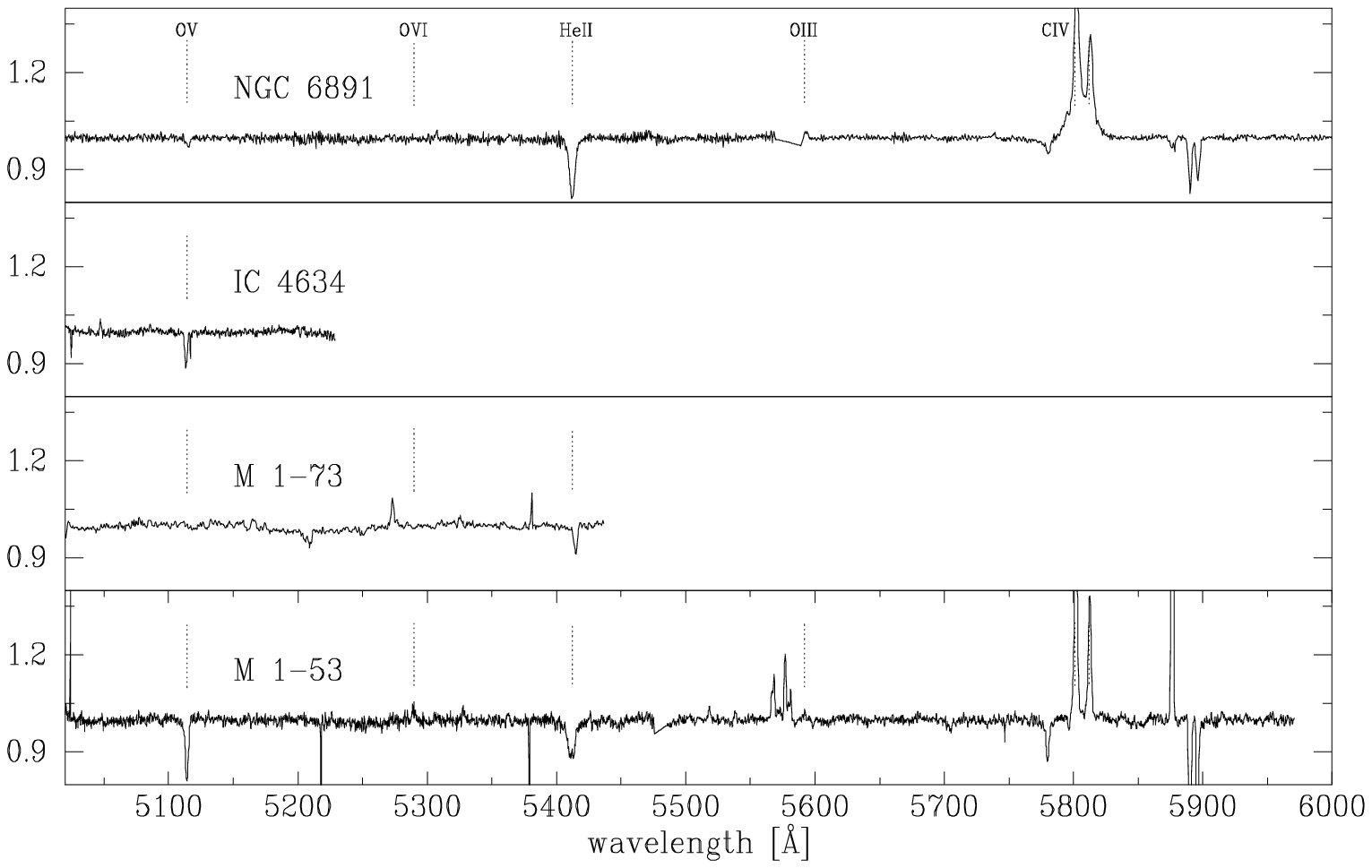}
   \includegraphics[width=0.99\textwidth]{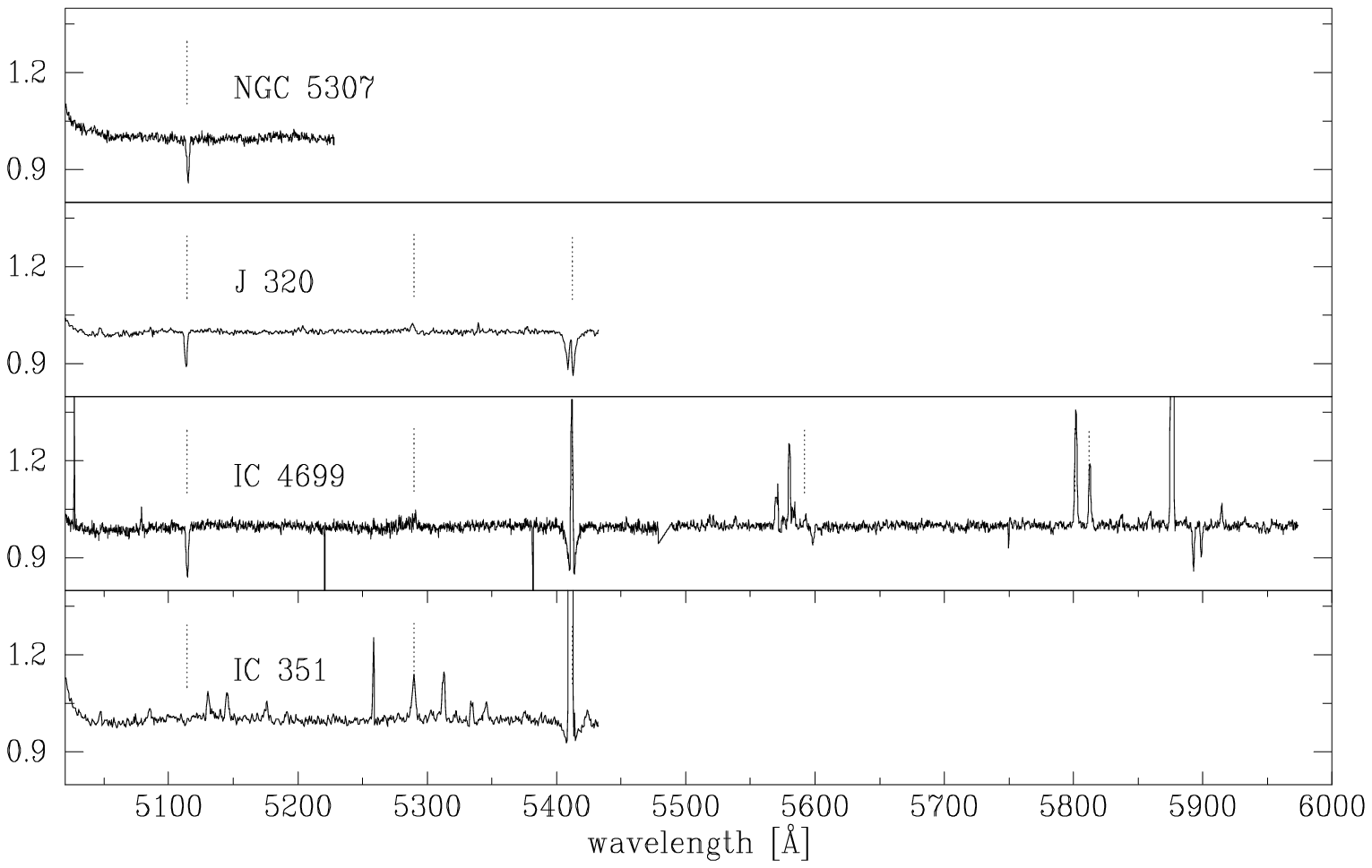}
      \caption[]{The ``red'' part of the spectra shown in 
                 Fig.~\ref{ST-type-0} to \ref{ST-type-3}.
                 Note the presence of the ions
\ion{O}{v} at 5114\AA,
\ion{O}{vi} at 5290\AA,
\ion{O}{iii} at 5592\AA,
\ion{He}{ii} at 5412\AA \ and
\ion{C}{iv} at 5806\AA.
}
         \label{ST-red1}
   \end{figure*}

\begin{figure*}
   \centering
   \includegraphics[width=0.99\textwidth]{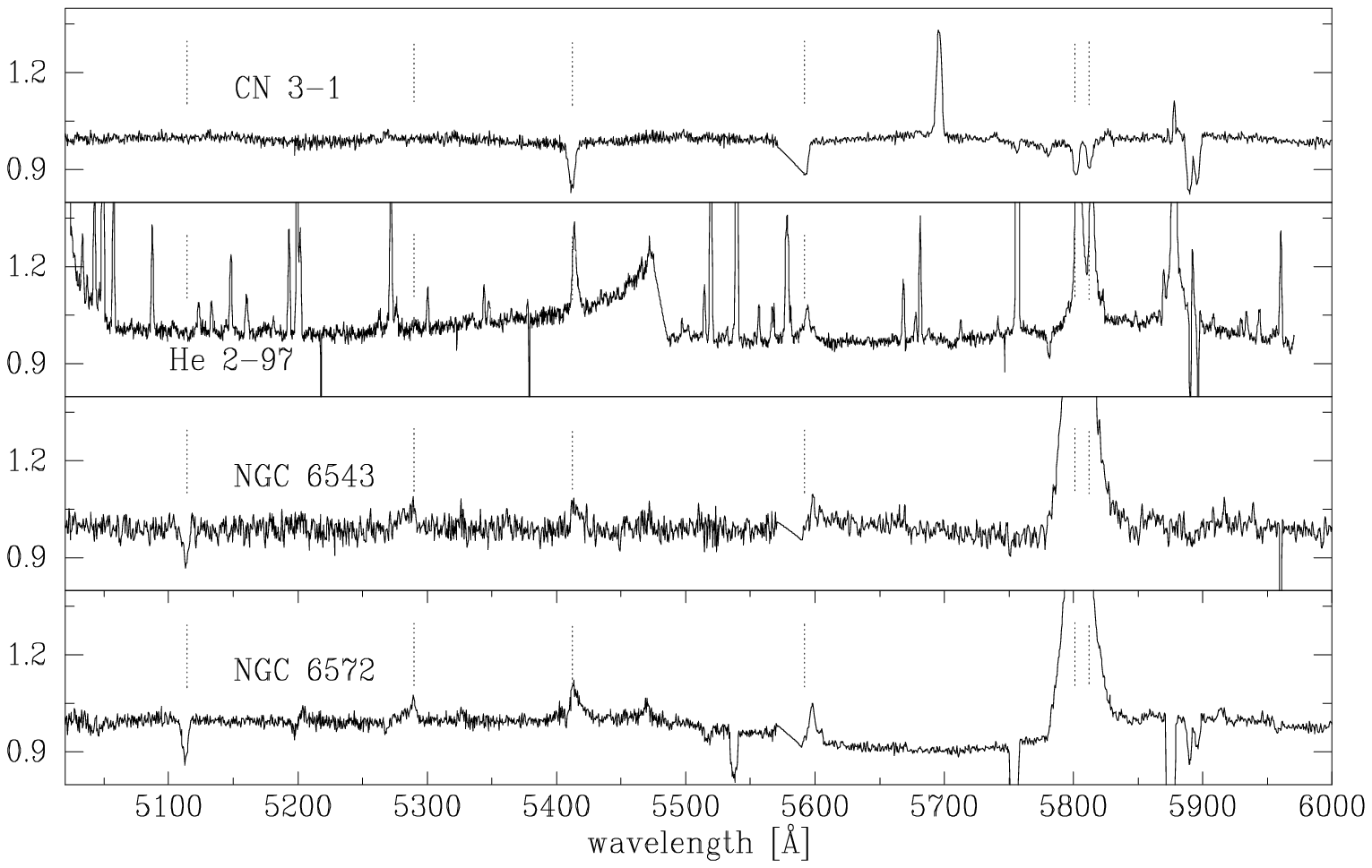}
   \includegraphics[width=0.99\textwidth]{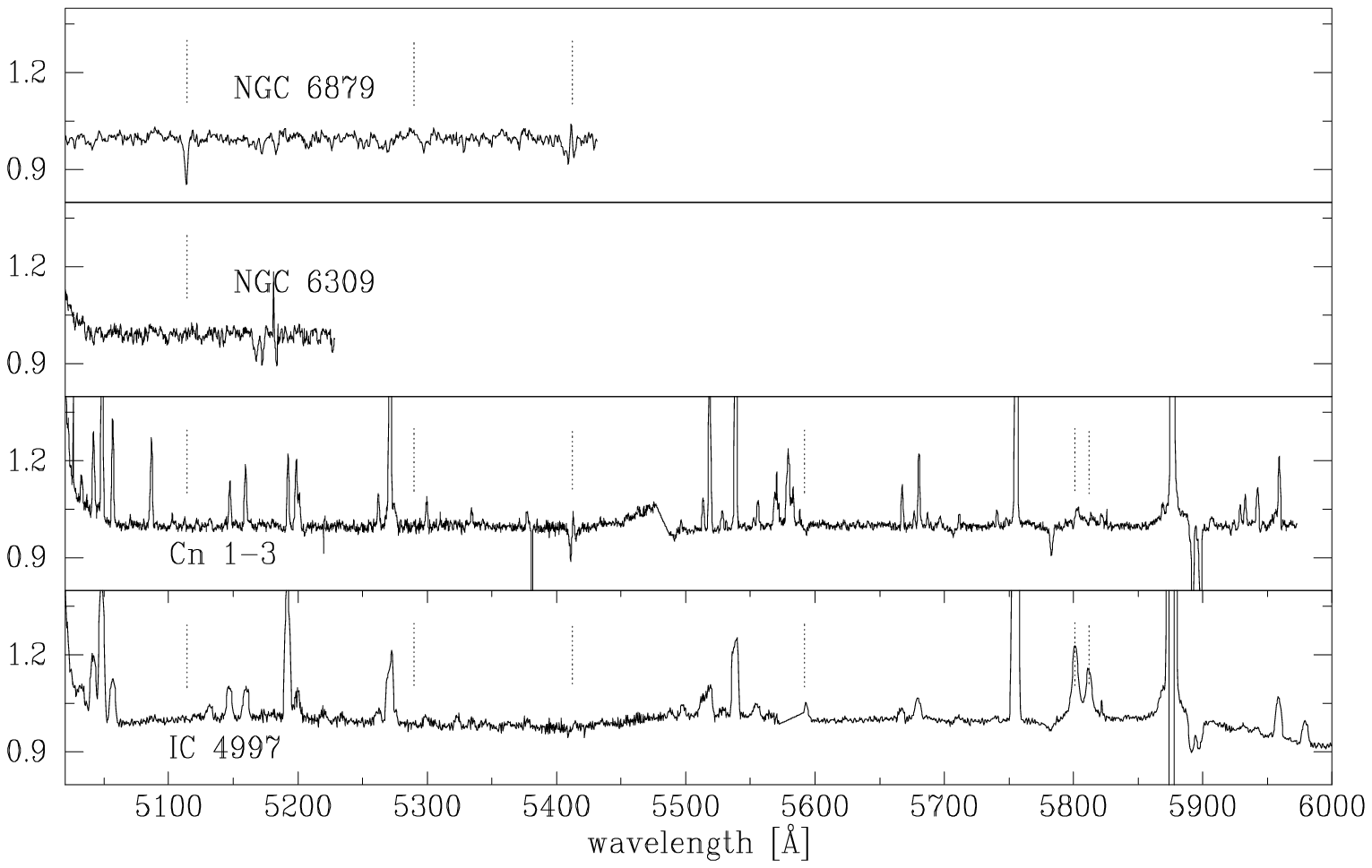}
      \caption[]{Continuation of Fig.~\ref{ST-red1}.}
         \label{ST-red2}
   \end{figure*}

\begin{figure*}
   \centering
   \includegraphics[width=0.99\textwidth]{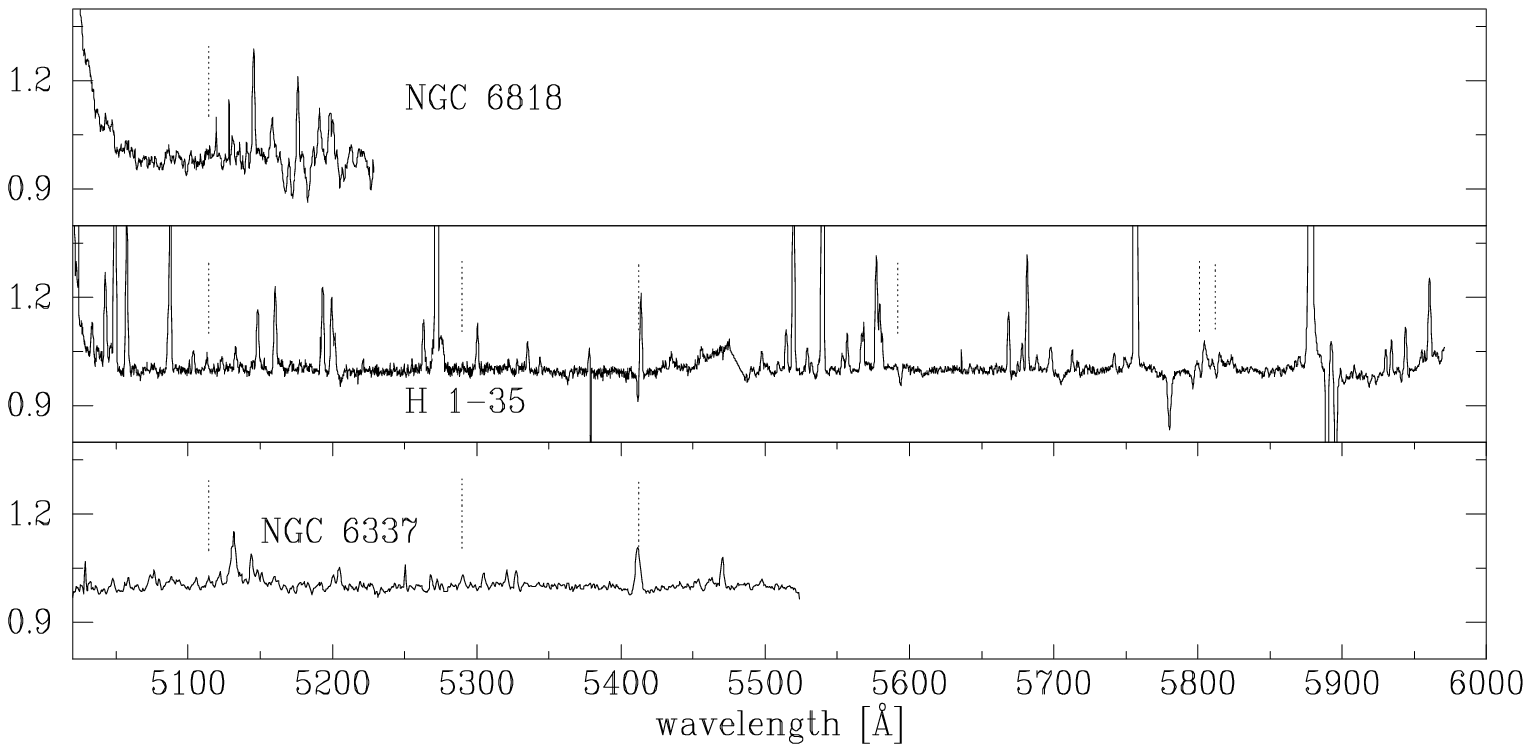}
      \caption[]{Continuation of Fig.~\ref{ST-red1}.}
         \label{ST-red3}
   \end{figure*}


\begin{acknowledgements}
We thank the referee Dr. Klaus Werner whose very useful remarks helped us to improve this paper.
This work was primarily based on observations 
obtained at the Gemini Observatory, which is operated 
by the Association of Universities for Research in 
Astronomy, Inc., under a cooperative agreement with 
the NSF on behalf of the Gemini partnership: the 
National Science Foundation (United States), the 
Science and Technology Facilities Council (United 
Kingdom), the National Research Council (Canada), 
CONICYT (Chile), the Australian Research Council 
(Australia), Minist\'erio da Ci\^encia e Tecnologia 
(Brazil) and Ministerio de Ciencia, Tecnolog\'ia e 
Innovaci\'on Productiva (Argentina). 
\end{acknowledgements}


\bibliographystyle{aa}
\bibliography{aa3}

\end{document}